# Navigation Assistance and Web Accessibility Helper

**Abdelhakim Herrouz[1], Chabane Khentout[2], Mahieddine Djoudi[3]**

[1]Department of Computer Science,
University Kasdi Merbah of Ouargla, Algeria
*abdelhakim.herrouz@gmail.com*

[2]Laboratoire des Réseaux et des Systèmes Distribués,
University Ferhat Abbas of Sétif, Algeria
*khentout@yahoo.fr*

[3]Department XLIM-SIC UMR CNRS 7252 & TechNE Research Group,
University of Poitiers
Téléport 2, Boulevard Marie et Pierre Curie, BP 30179
86960 Futuroscope Cedex (France)
*mahieddine.djoudi@univ-poitiers.fr*

### ABSTRACT

*Web accessibility is actually the most important aspect for providing access to information and interaction for people with disabilities. However, it seems that the ability of users with disabilities to navigate over the Web is not dependent on the graphical complexity, but on the markup used to create the structure of the website. Consequently, it's necessary to design some software assistants to help all users to mark themselves in space during a navigation session. In this paper, we propose an assistant for browsing on the Internet to allow user to get one's bearings within Web navigation. We describe our approach which puts at the disposal of the user a visited site map, thus giving an explicit representation of virtual space. Different levels of visualization are implemented in order to make the map more visible and less overloaded.*

### 1. INTRODUCTION

Presently, the Web has become one of the most widespread platforms for information change and retrieval. The web is an open space of information, dynamic, distributed, heterogeneous and not moderated.  On the Web, sites appear and disappear, content is modified and it becomes impossible to master their organization [5]. The availability of browsers for multiple computing platforms, many of them distributed for no cost, combined with new avenues for accessing the Internet allows even novice computer users with limited resources to make use of the wide range of services and information available on this global computer network.

To help users orient themselves in a hypertext Web, browsers often provide a list of the documents a user has visited, a way to move forward and backward along previously traversed links, and a quick way to return to a home document. These navigation aids are essential in helping users manage the huge store of information available on the Web. Hypertext links encourage users to explore related topics and references to other works from within a document. Although the backtracking aids and history list are helpful navigation tools, users often have trouble revisiting a page that was previously viewed in a session. This problem becomes acute after many invocations of the backtracking shortcuts. Users of hypertext systems often find themselves eagerly following hypertext links deeper and deeper into a hypertext Web, only to find themselves "lost" in the sense that they are unable find their way back to previously visited pages. This difficulty in revisiting previously viewed pages may discourage users from engaging in such exploratory behavior. It is hoped that the addition of the graphic history view will encourage exploratory behavior and help users navigate the Web more easily.

### 2. WEB ACCESSIBILITY

Web accessibility fundamentally means that people with disabilities (low vision, motor disabilities, or cognitive disabilities) can use the Web. It means that people with disabilities can perceive, understand, navigate, and interact with the Web, and that they can contribute to the Web. Web accessibility also benefits people without disabilities in certain situations, such as people using a slow Internet connection, people with "temporary disabilities" such as a broken arm, and people with changing abilities due to aging (visual, hearing, etc.) [20].

## 3. NAVIGATION PROCESS DIFFICULTIES

The Web combines difficulties that are usually present whenever a huge information system is used, with conceptual difficulties linked to the choices and the progression through heterogeneous information [4]. The difficulties encountered during navigation are various but they can be classified into two general types: the disorientation and the cognitive overhead [19].

Disorientation [2] can be defined as the mental state of feeling lost when navigating in hypertext systems. It is a psychological state resulting from problems in constructing pathways across a hypertext. The indications of disorientation based on the self-reported research data show that users:

(1) do not know where to go next;
(2) know where to go but not how to get there;
(3) and, do not know where they are in relation to the overall structure of the document.

Consequently, they may become frustrated, lose interest, and experience a measurable decline in efficiency.

The cognitive overhead happens with a user who has only a screen to work with. This user has to know the information shown is associated with what. Many decisions have to be taken while going through a hypermedia: which link to follow, how to retrieve the ones that are of interest among the links already visited or to be visited.

The user should be able to find the information being searched while moving from one page to another by following the different links. These tasks of searching for what is needed require accessing the information in smart way. This means that we need to have the capabilities to go from one place to another, identify the document reached, evaluate it, to save it or memorize its address, and related to other documents and information.

It is very common to notice that during the use of hypermedia, the user, after few minutes of search, does not know where he really is with respect to the different notions he went through. We reach a point where we start to move from one page to another or from one site to another without gaining anything new even if some of pages and/or site may contain relevant information. This is not going to improve the knowledge of the learner [15].

Working with World Wide Web may lead the user, from one link to another, to a page that has very little to do with the subject being searched for. The information read that is not related to a specific cognitive project is forgotten very quickly. Meanwhile we forget other pages that we have consulted earlier which contained information that is of interest to us. We activated a link that we taught it would allow us to get more information about the topic. This action took us further away from the subject because we kept following other links. Before we noticed it, we lost track the pages that interest us [13], [14]. After a half-hour of search, we turn off our computer with the impression that we went through a lot of material without learning anything new.

## 4. NAVIGATION HELP

Navigation help can be of two different ways: The first way is concerned with the construction of websites. A construction method should be adapted to make it easy for the user to access and search the sites. In [17] for example, the author proposes to limit the depth decomposition of the page to four levels. This means that only three nodes can be active at the same time. In addition, each screen should have about five active links. In order to be clear and efficient links to general ideas of dependant information are favored. This approach of construction will result into hypermedia with a simple structure, which is more efficient. The inconvenience of this method is that the user has to split for example a design of complete course into subsections, which are accessed separately. But we can always links these subsections to each other indirectly.

The second way is to provide a set of computer-aided tools that will allow the client user to navigate the Web easily using his/her preferred browser. The general browsers, Netscape or Internet Explorer propose some functionality such as history, and bookmarks but these kinds of help are insufficient for the user needs. In addition, the users of a hypertext system create different representations.

Many computer-aided systems to help the users to browse the Internet have been proposed in the literature. Among them are Broadway [6], FootPrints [24], Hypercase [11], Letizia [10], Mawa [18] and Nestor [25]. A comparative study of some of these tools is available in [4]. Nestor and Broadway are the closest to our design of computer-aided tools to navigate the Internet.

NESTOR [25] was developed at CNRS-GATE laboratory. It is a Web browser that draws interactive web-maps of the visited Web space during navigation: the objects that show on Nestor maps are the visited Web documents and the links that have been used to reach them. The web-maps are hybrid in the sense that users can add objects of their own – concepts, links, personal documents, organizers – and progressively evolve the maps into concept-maps. The maps are interactive in the sense that they provide direct navigation back to the represented objects, and allow for a full set of drag-and-drop operations aimed at structuring the information extracted from the Web: Nestor combines graphical Web navigation and mind-mapping features. Nestor is also collaborative software that enables small groups of people to share their navigation experience. We could say that Nestor promotes a constructionist approach to Web information mapping. This navigator is built to achieve the following two main goals: help the trainee to become an active leaner and make the browsing easy because most of the users have little experience with Internet. It is important to help them make full use of their experience [23]. Nestor is a complete and excellent navigator. It is a very good tool to build the navigation map. However, the client software is platform dependent; it runs only on top of Microsoft Internet Explorer on Microsoft Windows platforms.

The navigation helper Broadway (a BROwsing ADviser reusing path WAYs) is a server that keeps track of document requests made by the customers by saving them [6]. Broadway can be accessed by a group of users and supports indirect cooperation. It uses a reasoning system based on cases to advise a group of users on the interesting pages to visit according to the path that the group has already traversed. It establishes the reasoning system from cases that confirm to a flexible and generic framework formed by an index model of different situations. It helps a user who is navigating on the Web and facilitates the task of searching information on this hypermedia. The interaction of the user with Broadway is assured by the assistance of two means: the tool bars and the controller. A detailed and flexible behavior management is possible due to the extensive observation combined with the indexing model [19]. Broadway does not include the navigation time as a user parameter. But it remains a very good tool to model the user behavior during a browsing session and it has an open and well-adapted architecture to the Web.

## 5. SOFTWARE ARCHITECTURE

In order to allow the user to keep track of time and to know where he/she is, we have designed and implemented a computer-aided system for navigation on the Web [5]. This system, which was developed using Delphi, can be used with any browser (Firefox, Microsoft Internet Explorer or other). The main screen is made up of many windows. Its kernel is made up of two important modules: one is to collect the different URL addresses and the other is to build and interact with the graphical map and the management of navigation time [9], [3]. The user has access to a dictionary containing the frequently used words in Internet that may not be understood. Also help for the system can display in a separate window. This tool is designed to satisfy guidelines of accessibility of the W3C recommendation for disabled authors and learners especially with mobility impairments.

In order to guarantee that our system is independent of the browser, the way we recuperate the addresses of the sites/pages visited is using a proxy server. This proxy server seats in between Web clients and information servers using different protocols (Figure 1). It is used to pass the information from one end to the other. The client sends each user's request to the proxy server, which will respond directly if it has the information in its cache, or it will pass the request to the destination server. The proxy server keeps a copy of each document it sends in its cache. This copy is kept for variable amount of time. This way if a document is requested and is available in the cache of the proxy there is no need to get it from the destination server. The management of the cache is done based on the following parameters: date of the last time when the document was updated, maximum time that a document can spend in the cache and for how long has the document been in the cache without being used. This service, which is transparent to the user, makes the responses to the user requests more efficient. It also reduces the traffic on the network.

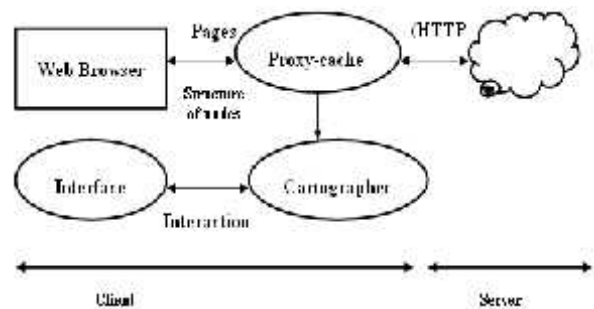

**Figure 1** Software Architecture

The proxy server receives the requests from the browser, rearrange them if needed and sent them to the module that is responsible to build the map. This server is installed locally on the user's machine to serve as a link for HTTP requests. The browser has to be configured to use this proxy server. Each HTTP request will be intercepted and sent by the proxy after extracting the necessary information (address requested, elapsed time since the last time this address was requested) and saves it. This data is stored in a file that will be used by the module responsible for building the map later on.

## 6. GRAPHICAL MAP FOR NAVIGATION

The development of a graphical map and its use as a computer-aided tool for Web browsing is based on the studies of cognitive processes that happen during the navigation of distributed hypermedia. It is a graphical representation at the same time of conceptual and geographical search path followed by a user while searching for a particular topic. The Navigation map that we designed is based on the idea used in conceptual maps [6].

A conceptual map is a way of representing the relationship between a set of knowledge and the nature of this relationship. It is a graphical representation of links among different concepts about the same topic. It should evolve with the knowledge of the trainee.

The conceptual map is also a computer-aided tool for navigation. It allows a hypertext reader to see on the screen the titles of information units and the links that connect them in a form of a network. It is drawn with a goal in mind, within well-defined references, and according to a graphical representation suitable for browsing problem.

### 6.1 Graphical Representations

Browsing the Web implies the manipulation of huge amount of information. The major role of the graphical interface of system developed for this purpose is to make this information easy to comprehend by the users. This is based mainly on the graphical representation of the different pieces of information and the relations connecting these pieces together. The graphical interface between the users and the system is a way to construct the image of the system. A review of the literature indicates the existence of many graphical representations.

The taxonomy developed in [22] is based on the notion of the user's actions. The classification proposed emphasizes the nature of actions (direct or indirect selections), their levels (single, group, and attributes and objects integrity) and their effect on the graph, on the representation and the transformation or organization of the objects selected.

The study proposed by [7] classifies representation techniques in five categories: geometric, network based, hierarchy, pixel oriented, and iconic. This approach has the disadvantage of mixing construction and graphical tools used as classification criteria, which makes it very difficult to characterize some systems.

The approach described in [17] is based on the type of data represented and the low level task performed by the user on this data. The author then listed different graphical representations used for each type of data. He also identifies seven task types that the graphical representation should favor. The high level tasks that are independent of the data being manipulated are: general view of the information, zooming, filtering, getting the details, link representation together, having a history of actions performed, and extracting part of the information so that it can be used by other applications. Three of these points (general view of the information, zooming, and getting the details) are considered during the conception of the representation.

In [1], the authors propose to characterize the graphical representation based on a chosen point of view about the data but not on the type of data. A point of view is defined by deciding what is necessary out of the data that should be given to the users based on his needs to perform his task in a satisfactory manner. If we are unable to characterize in a precise way the object's activities then the graphical representation should be flexible enough to detect one or many points of view that are suitable to accomplish the task. For a set of data we might have more than one point of view depending on how the data is considered. These points of views might complement each other for the purpose of the user's activities. So it necessary to be able to represent simultaneously many views which means we should choose a graphical representation guided by multiple points of view. This corresponds to multiple views discussed in [12] and [25]. This multiplicity should be taken as a factor during the design of an interface that can adapt itself to different tasks.

### 6.2 Choice of Graphical Representation

The navigation map gives the possibility to keep track of path followed by the user while browsing the Web. The map is modeled by a directed graph. Each page address (URL), the topic or title of the page, and the time spent connected to this page are kept in the nodes. The map is displayed upon request of the user at any stage of the browsing.

A directed graph representation of the map is most suitable for its visualization. Each node contains the name and the information of the page visited. The information kept should be in such way that it does not affect the clarity of the graph. The nodes are connected to each other to indicate the fact that the user has moved from a specific page to another. The nodes should be displayed on the screen in a way that all are visible and with minimum edges intersection.

To choose the best representation of the map, we looked into different techniques (available in the literature) to display graphs. Also we kept in mind the specific properties of our graph and the different operation that are performed on it. We found that the tree representation is the most suitable for our case. In this approach, all the nodes are drawn. The information represented in the graph is very easy to read. The user can modify this representation as it is explained later. The system allows the user to display different information as a part of the tree. The user can selectively display document titles, URLs, or a thumbnail image for each node. When the mouse is placed over a node in the tree, the title and URL of the document appears near the mouse. A user can recall a document in the tree by double-clicking on a node in the tool window.

### 6.3 Manipulation of the map

In addition to the automatic graphical map generation representing the visited pages, the system allows the drawing of the map from a list of identifiers of pre-selected pages. Also the user can follow the map evolution by creation, deletion of any link or reorganization of the graph, or do only a read of the map for a simple task. All actions that are performed and amount time spent on each page are saved and used for evaluation. This information can be shared among a group of users.

The user has also the possibility to save, print or reopen the map constructed during a navigation session. He has also access to the log report during a session that will allow him to do a self-evaluation and to be able to follow his progress during a training period or a search for information. It is also possible to have report indicating the daily interactions and the time spent connected to each site (Figure 2 and Figure 3). The graphical map can be used to share information within a group of learners in a cooperative learning environment [6], [8]. Each user can benefit from the experience of the other members of the group [21].

### 7. EXPERIMENTS

These tools presented here were in response to certain browsing problems. There are a number of assistants tools that give the possibility to the users to experiment. We concentrated on the problem of disorientation and putting some reference points. During the experiment, the use of usual browsers, such as Internet Explorer or Firefox, by novice users while solving a pedagogic task, is observed.

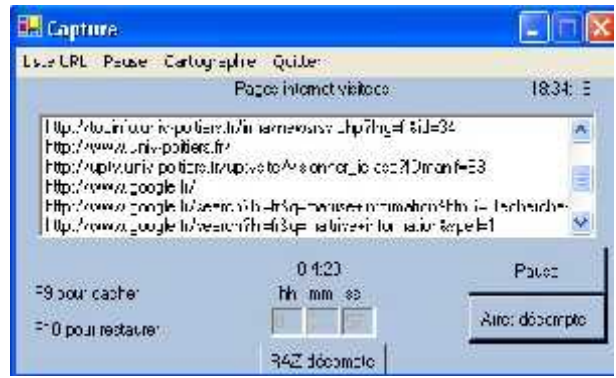

**Figure 2** Display of URL List

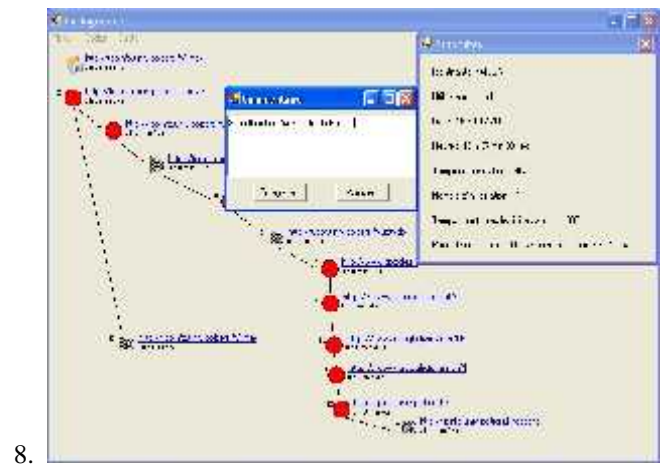

8.

**Figure 3** Display of the Map and Time Spent

We have tested the set of tools developed in a real practical sessions. There were about 100 users, aged from 19 to 22 years. All had some experience with some browser. They used the Internet to search for information before. This is a limited experience because we have only the strict minimum needed equipment. Also the connection equipment is not suitable for heavy use.

The experimental environment is made of: free access to the Web. A guided access to the course according to a plan, prepared by the supervisor, which is made of a set of documents on the educational server and some links to public documents available on the Web.

The collection of information about a particular topic from the Internet and the structure of this information into a personnel or group document will be submitted to the teacher using the browsing map.

The proxy architecture made it possible, while using the tools, to display on the screen the browser on one window, the sequence of site and the navigation map on another. This solution helps to reduce the cognitive overload of the users.

For the supervisor the graphical map can be considered as a tool to analyze the content of what is being taught, to have a better structure of the programs and manuals, and to build a plan of the course. The preparation of a guided tour with comments helps to get the new user to start. These guided tours allow a simple browsing without limiting the freedom of exploring. They include some public documents available over the Internet and some local documents prepared for pedagogical purpose.

### 8. CONCLUSION

Our implemented Web-client can be considered as constituting an add-on to standard Web browsers. It presents a way to solve many navigation and Web accessibility problems. The system helps people with disabilities to use and contribute to the Web. However, we envisage to evaluating the system in a Web accessibility environment. This will allow us to measure the success of our tool in simplifying the navigation procedures for people with disabilities. Also, we plan that users will be able to annotate nodes with personal notes, and describe their preferred path through the information space.

## References

[1] Bruley C. And Genoud, P.  Contribution à une taxonomie des représentations graphiques de l'information. Dixièmes journées francophones sur l'Interaction Homme Machine, IHM 98, Nantes, 1998.
[2] Cangoz, B., Altun, A. The effects of Hypertext Structure, Presentation and Instruction Types on Perceived Disorientation and Recall Performances. In Contemporary Educational Technology, Volume 3, Issue 2, pp. 81-98, 2012.
[3] Djoudi, M. Conception d'assistants à la navigation sur l'internet. Dans Comprendre les usages de l'internet. d'Eric Guichard Edition ENS Rue d'ULM, ISBN 2-7288-0268-8, 2001.
[4] Herrouz, A., Khentout, C., Djoudi, M. Overview of Visualization Tools for Web Browser History Data. IJCSI International Journal of Computer Science Issues, Vol.9, Issue 6, No3, November 2012, pp. 92-98, 2012.
[5] Herrouz, A., Djoudi, M. Conception d'un Système d'Assistance à la Navigation et à l'Apprentissage sur Internet. in Seventh Maghrebian Conference on Computer Sciences (7th MCSEAI), Tome 1, pp. 59-76, Annaba, 2002.
[6] Jaczynski, M., B. Trousse. WWW assisted browsing by reusing pas navigations of a group of users. In Proceedings of the European Workshop of Case-base Reasoning, EWCBR'98, LNCS/AI, Dublin, Ireland, Spring-Verlag, 1998.
[7] Keim, D. A. Visual techniques for exploring databases. In Invited Tutorial, Int. Conference on Knowledge Discovery in Databases KDD'97, Newport Beach, 1997.
[8] Khentout, C., M. Djoudi and L. Douidi. "Roundup of Graphical Navigation Helpers on the Web", *Journal of Computer Science, ISSN: 1549-3636*, Vol. 3, No. 3, pp 154-161, 2007.
[9] Khentout C. Interfaces et assistance à l'apprentissage dans une université virtuelle. Thèse de Doctorat en Informatique, Université Ferhat Abbas, Sétif, Algérie, 2006.
[10] Lieberman, H. Letizia: An Agent that Assists Web Browsing. In Proceedings of International Joint Conference on Artificial Intelligence (IJCAI'95), pages 924-929, Morgan Kaufmann, 1995.
[11] Micarelli, A. and F. Sciarrone. A Case-Based System for Adaptive Hypermedia Navigation. In Advances in Case-Based Reasoning, Proc. of the 3rd European Workshop on Case-Based Reasoning (EWCBR'96), vol. 1168, pp: 266-279, Springer, 1996.
[12] Nigay L., Vernier F. " Design method of interaction techniques for large information spaces ", In Proceedings of Advanced Visual Interfaces, AVI'98, p. 37-46, 1998.
[13] Perriault J. Le temps dans la construction des savoirs à l'étude des médias. Revue européenne des sciences sociales, Tome XXXVI, 111 : 109-118, 1998.
[14] Perriault J. Synchronous and asynchronous media in an hybrid learning process: effects of time compression and expansion. European Distance Education Network (EDEN). In Proceedings of the Conference, Milton Keynes, The Open University, 1996.
[15] Quarteroni, P. Un hypermédia pédagogiquement efficace. Revue éducatechnologiques, sous la direction de J. Rhéaume, Université Laval, Canada, 1996.
[16] Rhéaume, J. Les hypertextes et les hypermédias. Revue éducatechnologiques, Faculté des sciences de l'éducation, Université Laval, Canada, 1997.
[17] Shneiderman B. Designing the User Interface. Addison Wesley, third edition, 1998.
[18] Singer, N. and  S. Trouillet. Cartographie du Web et navigation sociale : le système multi-agent MAWA. Conférence Nimes TIC : Interaction et Composition Dynamique, 2001.
[19] Souza, A.P. and P. Dias. Analysis of Hypermedia browsing processes in Order to Reduce Disorientation. In Proceedings of ED-MEDIA'96 conference, AACE, 1996.
[20] Talhi, S., Khadraoui, F., Djoudi, M. "Implementing WAI Authoring Tool Accessibility Guidelines in Developing Adaptive Elearning". *International Journal of Modern Education and Computer Science (IJMECS), ISSN: 2075-0161*, Vol. 4, No. 9, September, pp.1-13, 2012.

[21] Trousse B., M. Jaczynski and R. Kanawati. Une approche fondée sur le raisonnement à partir de cas pour l'aide à la navigation dans un hypermédia. In Proceedings of Hypertext & Hypermedia: Products, Tools and Methods (H2PTM'99). Paris, 1999.

[22] Tweedie L. A. Interactive visualization artefacts: how can abstractions inform design? Proceedings of the CHI'95 Conference, 247-265, 1995.

[23] Wang-Baldonado, M. Q. and T. Winograd. Sense maker: An information exploration interface supporting the contextual evolution of a user's interests. In Proceedings of ACM CHI 97 Conference on Human Factors in Computing Systems: 11-18, 1997.

[24] Wexelblat, A. Maes, P. Footprints: History-Rich Tools for Information Foraging. Proceedings of CHI'99 Conference on Human Factors in Computing Systems, ACM Press, 1999.

[25] Zeiliger, R., Esnault, L. The constructivist mapping of internet information at work with Nestor. In A. okada, S.B. Shum & T. Sherborne (Eds.). Knowledge Cartography, pp. 89-111, Springer, 2008.